\documentclass[conference]{IEEEtran}
\usepackage{cite}

\ifCLASSINFOpdf
  \usepackage[pdftex]{graphicx}
\else
\fi
\usepackage{amsmath}
\usepackage{algorithmic}

\usepackage{caption}
\usepackage{subcaption}

\usepackage{url}

\usepackage{color}
\usepackage{balance}

\begin{document}
%
\title{Optimized Ensemble Model Towards Secured Industrial IoT Devices}

\author{
\IEEEauthorblockN{MohammadNoor Injadat\IEEEauthorrefmark{1}} 
\IEEEauthorblockA{\IEEEauthorrefmark{1}Data Science and Artificial Intelligence Department, Faculty of Information Technology, Zarqa University, Zarqa, Jordan\\ 
e-mail: minjadat@zu.edu.jo}}


%


\maketitle

\begin{abstract}
The continued growth in the deployment of Internet-of-Things (IoT) devices has been fueled by the increased connectivity demand, particularly in industrial environments. However, this has led to an increase in the number of network related attacks due to the increased number of potential attack surfaces. 
Industrial IoT (IIoT) devices are prone to various network related attacks that can have severe consequences on the manufacturing process as well as on the safety of the workers in the manufacturing plant. 
One promising solution that has emerged in recent years for attack detection is Machine learning (ML). More specifically, ensemble learning models have shown great promise in improving the performance of the underlying ML models. Accordingly, this paper proposes a framework based on the combined use of Bayesian Optimization-Gaussian Process (BO-GP) with an ensemble tree-based learning model to improve the performance of intrusion and attack detection in IIoT environments. The proposed framework's performance is evaluated using the Windows 10 dataset collected by the Cyber Range and IoT labs at University of New South Wales. Experimental results illustrate the improvement in detection accuracy, precision, and F-score when compared to standard tree and ensemble tree models. 
\end{abstract}


%
\IEEEpeerreviewmaketitle

\section{Introduction}\label{Intro}
\indent The continued growth in the deployment of Internet-of-Things (IoT) devices has been fueled by the increased connectivity demand. Accordingly, it has been projected that the number of connected devices will reach around 28.5 billion devices by 2022 \cite{Cisco_data_growth}. This includes devices used in various applications such as healthcare \cite{IoT_healthcare}, smart cities \cite{IoT_smart_cities}, and intelligent transportation systems \cite{IoT_ITS}.\\
\indent However, this has led to an increase in the number of network related attacks. This is attributed to the increased number of potential attack surfaces with Forbes recently reporting that there were more than 2.9 billion events in 2019 \cite{IoT_attack_statistics}. This represented an increase by three-folds compared to 2018 \cite{IoT_attack_statistics}. In a similar fashion, a report by Sonicwall stated that there was an increase of 215.7\% in IoT malware attacks between 2017 to 2018 \cite{IoT_attack_statistics1}. Among the most vulnerable IoT devices are the industrial IoT (IIoT) ones. This is attributed to the emergence of the automated manufacturing application as one of the 5G ultra-reliable low latency (URLLC) use case \cite{IIoT_security1,IIoT_security2}. IIoT devices are prone to various network related attacks such as denial-of-service, eavesdropping, malware, and data tampering \cite{IIoT_security2}. These can have severe consequences on the manufacturing process as well as on the safety of the workers in the manufacturing plant \cite{IIoT_security2}. Although researchers have proposed multiple potential mechanisms to protect such environments \cite{SDP2,IoT_attack_survey,nti2022stacknet,muthiya2020design}, there stills needs to have more effective IIoT attack detection mechanisms to address such concerns.\\
\indent One promising solution that has emerged in recent years is Machine learning (ML). ML paradigms represent suitable solution in IIoT environments. This paradigm is particularly promising given the abundant amount of data generated by IIoT devices. ML-based systems ``learn'' without explicit programming, thus improving the systems' flexibility and dynamic nature \cite{Moubayed_IDS_chapter,Moubayed_thesis}. Moreover, such techniques have proven to be effective and computationally efficient in a multitude of applications \cite{Moubayed_DNS2,Moubayed_DNS3,Injadat_Moubayed_survey,Injadat_thesis,Aburakhia1}. As such, they are a prime candidate to be deployed in IIoT devices as part of any intrusion detection mechanism for such environments. Moreover, it is important to optimize the parameters of these ML models as this can further improve the performance by more accurately and effectively detecting the attacks \cite{Li_HPO}. This is particularly important given the fact that most previous related works use the default parameters of the ML models proposed, reducing their impact.\\
\indent Therefore, this paper proposes an optimized ensemble learning model that can accurately detect intrusions and attacks on IIoT devices and systems. More specifically, this paper develops a framework based on the combined use of Bayesian Optimization-Gaussian Process (BO-GP) with an ensemble tree-based learning model (that can reduce the variance and bias \cite{ensemble_bagging,ensemble_boosting}) to improve the performance of intrusion and attack detection in such environments. As such, a summary of this work's main contributions is:
\begin{itemize}
	\item \textit{Investigate} the dataset's characteristics through principal component analysis.
	\item \textit{Propose} an optimized ensemble tree-based learning model for intrusion detection in IIoT devices.
	\item \textit{Evaluate} the proposed model's performance in comparison to other models from the literature. 
\end{itemize}

\indent The remainder of this paper is organized as follows: Section \ref{related_work} provides a brief summary of some of the previous works from the literature. Then, Section \ref{proposed_approach} presents the proposed optimized ensemble learning approach to protect IIoT devices and discusses its complexity. Section \ref{dataset_description} describes the dataset considered in this work. Then, Section \ref{results} discusses the experimental setup and the corresponding results. Lastly, Section \ref{conc} concludes the paper.
\section{Related Work}\label{related_work}
\indent With the growth in attack on Internet-based services, particularly in industrial environments, the security of such environments has become a prime concern. To that end, multiple researchers have proposed ML-based mechanisms and frameworks to detect and mitigate network attacks on IIoT networks and devices  \cite{Injadat_Moubayed_survey,Injadat_thesis}. In what follows, a brief overview of the ML-based solutions proposed in the literature for general network attack detection and IoT network attack detection is given.
\subsection{General ML-based Network Attack Detection Works:}
\indent As mentioned earlier, ML has been proposed as a viable potential solution for network intrusion detection and accordingly has been proposed in many previous works from the literature \cite{Injadat_IDS0,Injadat_BO,Injadat_IDS1,Li_IDS,Li_IDS2,Injadat_IDS2,Moubayed_ICM2}. An optimized ML-based framework for intrusion detection using Bayesian optimization was proposed by Injadat \textit{et al.} that achieved a better detection accuracy and reduced the false alarm rate \cite{Injadat_BO}. The authors further extended their work by proposing a multi-stage optimized framework that further reduced the complexity of the ML-based models while still maintaining the detection performance \cite{Injadat_IDS1}. On the other hand, Li \textit{et al.} proposed the use of tree-based models to detect intrusions in intelligent transportation systems with high detection accuracy and low false alarm rate \cite{Li_IDS}. The authors again extended the work by proposing a multi-tiered hybrid framework capable of detecting known and unknown vehicular network attacks using signature and anomaly-based intrusion detection systems \cite{Li_IDS2}. In contrast, Salo \textit{et al.} proposed combining ensemble feature selection with clustering-enabled classification to detect zero-day attacks \cite{Injadat_IDS2}. Similarly, Moubayed \textit{et al.} also combined information gain-based feature selection with genetic algorithm and random forest models to detect botnet attacks using DNS query information \cite{Moubayed_ICM2}. 
\subsection{ML-based IoT Network Attack Detection Works:}
\indent ML-based frameworks and mechanisms have also been proposed for industrial and regular IoT network intrusion attack detection \cite{IoT_security_scada,IoT_security_smart_homes,IoT_security_DRNN,Moubayed_ICM1}. Teixeira \textit{et al.} proposed using an ML-based framework to detect network attacks on the control system of a water storage tank \cite{IoT_security_scada}. Similarly, Anthi \textit{et al.} proposed an ML-based framework for IoT intrusion detection in smart homes \cite{IoT_security_smart_homes}. In contrast, Almiani \textit{et al.} proposed using deep recurrent neural networks to detect IoT intrusions in an effective manner \cite{IoT_security_DRNN}. On the other hand, Injadat \textit{et al.} proposed combining Bayesian optimization and decision trees to detect botnet attacks in IoT environments \cite{Moubayed_ICM1}.\\
\indent Despite the significant interest in proposing ML-based frameworks for IoT intrusion detection, most previous works typically use default parameters for the corresponding models. Moreover, they often neglect the class imbalance problem which is common in IoT network attack datasets. Therefore, to further enhance the detection performance, optimized models need to be considered. 
\begin{figure}[!t]
	\centering
	\includegraphics[scale=.45]{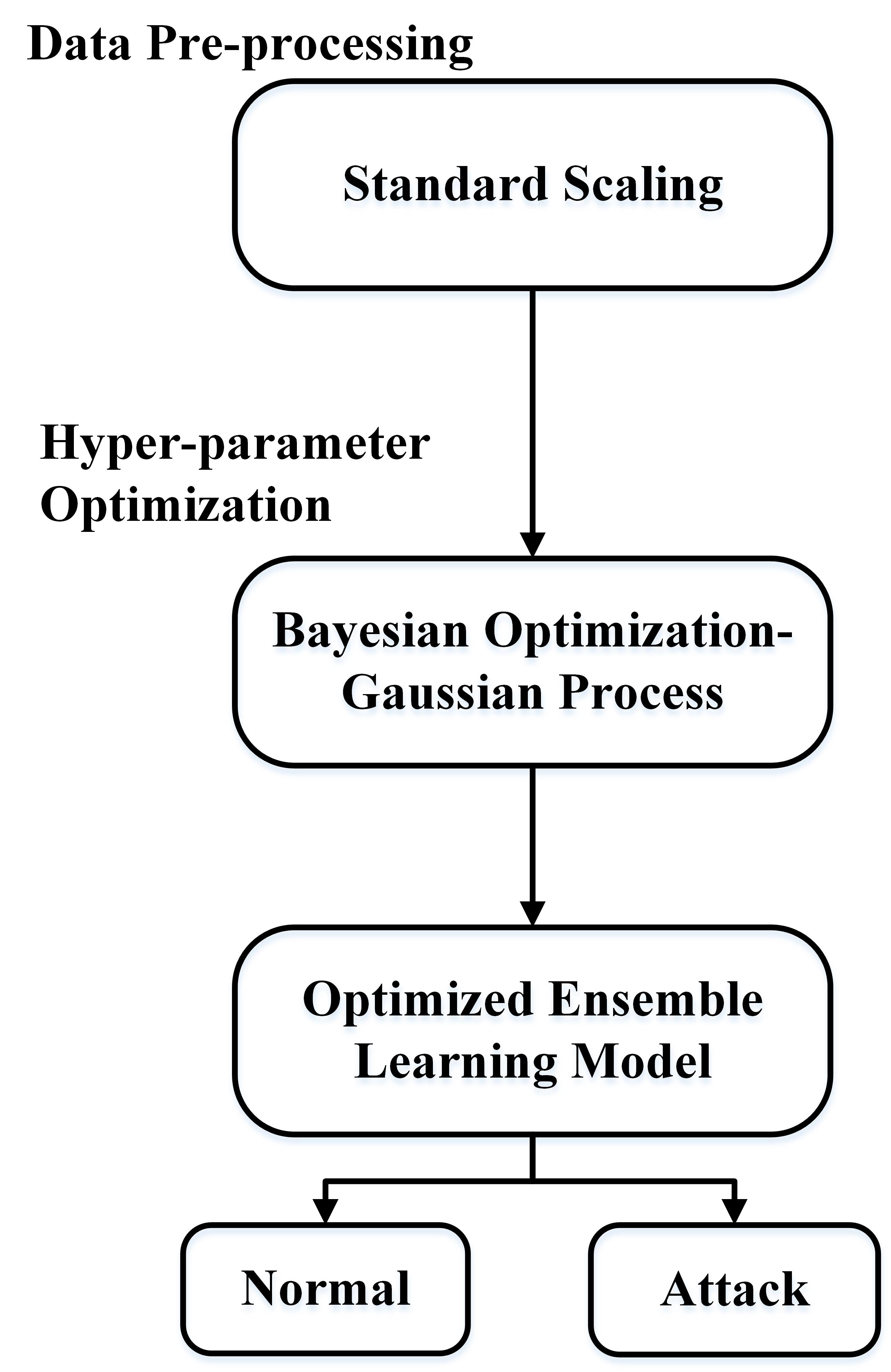}
	\caption{Proposed Optimized Ensemble Learning Framework for Intrusion Detection in IIoT Environments}
	\label{iot_approach_fig}
\end{figure}
\section{Proposed Approach}\label{proposed_approach}
\subsection{Description:}
\indent This paper proposes a combination of BO-GP with ensemble tree-based learning model to optimize the ensemble learning model's parameters. As illustrated in Fig. \ref{iot_approach_fig}, the proposed framework includes two main portions:
\begin{enumerate}
	\item Data pre-processing: This portion focuses on preparing the data in such a way that would maximize the ensemble learning model's performance. To that end, two steps are taken. The first is replacing any missing values with the mean value of the feature. The second is performing standard scaling using each feature's mean and standard deviation as per the following equation:
	\begin{equation}
		x_{scaled}=\frac{x-\mu_x}{\sigma_x}
	\end{equation}
	where $\mu_x$ is the mean vector and $\sigma_x$ is the standard deviation vector of the training samples. This is done to unify the scale of all features to be zero mean and unit standard deviation.
	\item Hyper-parameter Optimization: This portion focuses on optimizing the hyper-parameters of the ensemble learning model. This includes the type of ensemble to be used, the number of learners included in the ensemble, and the parameters of the learners used. Therefore, BO-GP algorithm is used to optimize the ensemble learning model's parameters. This algorithm is one of the probabilistic global optimization models commonly used for hyper-parameter optimization \cite{Injadat_IDS1}. It is worth noting that this algorithm is chosen due to its ability to determine near-optimal hyper-parameter combinations in a small number of iterations \cite{Li_HPO}.  
\end{enumerate}
Note that tree-based ensemble models are considered in this work since they have proven to be effective in intrusion detection frameworks \cite{Moubayed_DNS2,Moubayed_DNS3}.
\subsection{Computational Complexity:}
\indent To determine the overall complexity of the proposed framework, the complexity of each of its individual components is investigated. To do so, it is assumed that the dataset considered consists of $I$ training instances and $F$ features. The complexity of the standard scaling component is $O(F)$ since this process involves calculating the mean and standard deviation of each of the features. Next, the complexity of the BO-GP algorithm is considered to be $O(I^3)$ as shown in  
\cite{Injadat_IDS1}. Lastly, the complexity of the ensemble learning model depends on the base learners used to build it. As mentioned earlier, this work assumes that tree-based learners are used as the basis of the ensemble model since they provide highly accurate prediction and can deal with data that is either linear or non-linear \cite{Moubayed_DNS2,Moubayed_DNS3}. In general, the complexity of the ensemble tree-based learning model would be $O(I^2 F)$ \cite{Li_IDS}. However, since tree-based models as well as ensemble models can be trained in parallel, the complexity of the optimized ensemble learning model is reduced to $O(\frac{I^2 F}{T})$ where $T$ is the maximum number of participating threads in the training process \cite{Li_IDS}. Therefore, the overall computational complexity of the proposed optimized ensemble learning framework is considered to be $O(I^3)$. 
\section{Dataset Description}\label{dataset_description}
\begin{figure}[!ht]
	\centering
	\includegraphics[scale=.6]{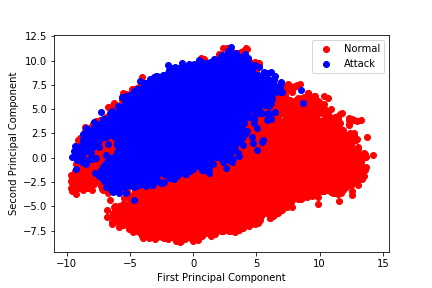}
	\caption{First and Second Principal Components of Windows 10 Dataset}
	\label{pca_iot}
\end{figure}

\indent The dataset used in this work is the Windows 10 dataset which is part of the Ton IoT datasets described in \cite{ton_dataset}. The Cyber Range and IoT Labs at the University of New South Wales Canberra collected this dataset using a realistic and large-scale network \cite{ton_dataset}. More specifically, this work focuses on the Windows 10 dataset since many industrial IoT devices are using it as an operating system \cite{IIoT_windows,IIoT_windows1,IIoT_windows2,IIoT_windows3}. In what follows, the statistics of the dataset are described briefly.\\
\indent The dataset is collected using multiple IIoT devices running Windows 10. The data is collected for both attack and normal operation duration with multiple attack types being initiated. Each data sample collected includes around 125 features corresponding to the process activity, the corresponding processor (CPU) and memory activity, network activity, and disk activity. Moreover, the respective label of the data sample (normal or attack) is also given. In total, the dataset consists of 24,871 \textbf{Normal} instances and 11,104 \textbf{Attack} instances. It can be seen from this description that the dataset does have a level of class imbalance. This needs to be considered when designing the corresponding intrusion detection framework. Note that no data augmentation process was conducted in this work to avoid any potential bias that may be introduced by the augmentation algorithm that is used. Further details about the dataset can be found in \cite{ton_dataset}.\\
\indent A principal component analysis is conducted with the first and second components being plotted in Fig. \ref{pca_iot}. It can be seen that the dataset is non-linear. This implies that the corresponding ML models developed need to be able to deal with such non-linearity without affecting the detection performance.
\section{Experiment Results \& Discussion}\label{results}
\subsection{Experiment Setup}
\indent \textbf{MATLAB 2022a} is used to evaluate the performance of the proposed framework. This is done using a laptop having Intel\textsuperscript{\textregistered} Core\textsuperscript{TM} i7-9750H CPU 6 Cores at 2.6 GHZ and 16GB of memory running Windows 10.  
\subsection{Performance Metrics}
\indent To evaluate the performance of the proposed optimized ensemble learning framework, four metrics are used, namely the detection accuracy, precision, recall, and F-score. This can be calculated using the following equations \cite{performance_metrics,Injadat_IDS1}:
\begin{equation}
	Accuracy = \frac{(TP + TN)}{(TP + TN + FP + FN)}
\end{equation}
\begin{equation}
	Precision = \frac{TP}{(TP + FP)}
\end{equation}
\begin{equation}
	Recall = \frac{TP}{(TP + FN)}
\end{equation}
\begin{equation}
	F-score = 2\times\frac{(Precision\times Recall)}{(Precision + Recall)}
\end{equation}
\begin{figure*}[!t]
     \begin{subfigure}{.5\textwidth}
         \centering
         \includegraphics[scale=.2]{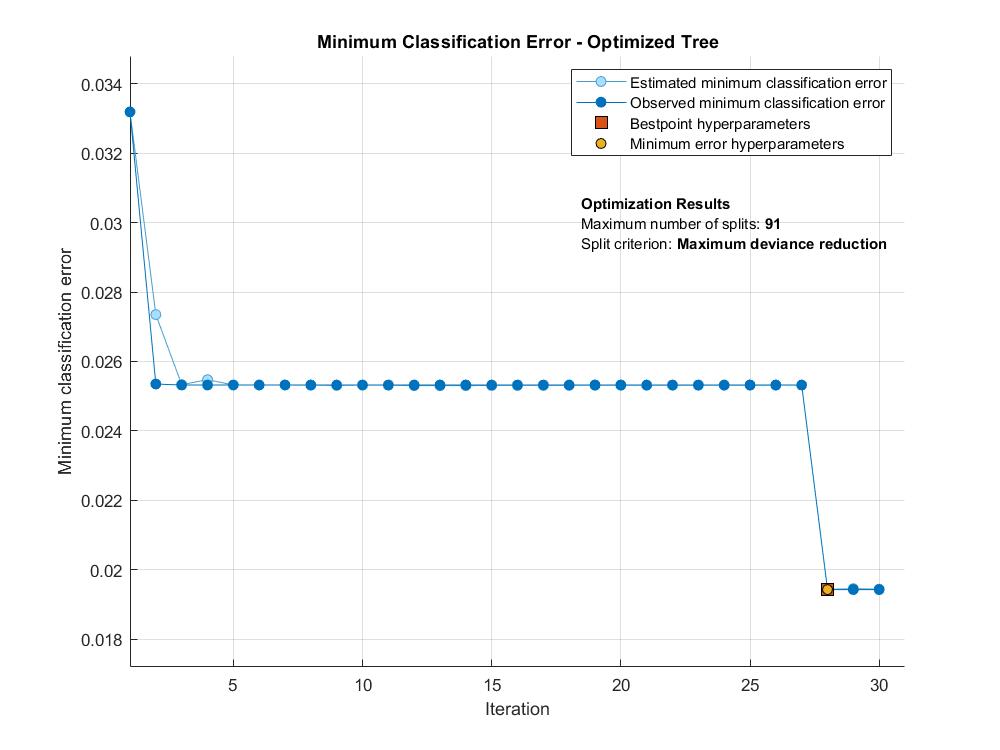}
         \caption{Minimum Classification Error Progression-Optimized Tree}
         \label{opt_tree_min_error}
     \end{subfigure}
     \hfill
     \begin{subfigure}{.5\textwidth}
         \centering
         \includegraphics[scale=.2]{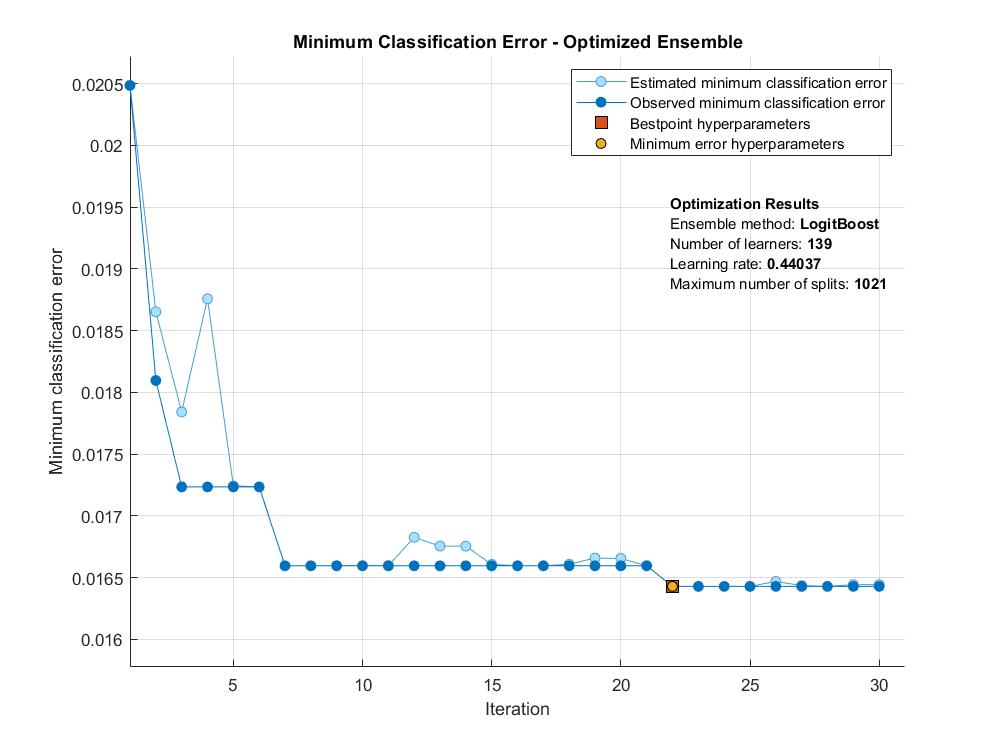}
         \caption{Minimum Classification Error Progression-Optimized Ensemble}
         \label{opt_ensemble_min_error}
     \end{subfigure}
        \caption{Minimum Classification Error Progression During Optimization}
        \label{Min_class_error}
\end{figure*}
\subsection{Proposed Framework's Performance Results \& Discussion}
\indent The performance of the proposed optimized ensemble tree-based learning model is compared to that of other ML-based models proposed in the literature (such as \cite{Moubayed_DNS2,Moubayed_DNS3}) including the standard decision tree model (denoted as DT), optimized decision tree model (denoted as Optimized DT), Bagging Ensemble Tree model, Boosting Ensemble Tree model, and Random Under-sampling Boosting Ensemble Tree model (denoted as RUSBoosting). The reason behind choosing these specific ensemble models is as follows. Starting with the bagging ensemble, it was chosen because it can reduce the classification variance of the based learner, improve its accuracy, and maintain its low bias \cite{ensemble_bagging}. For the boosting ensemble, it was chosen because it can reduce the classification bias and improve the classification accuracy \cite{ensemble_boosting}. Finally, the RUSboosting ensemble is chosen as it achieves the same goal as regular boosting while being able to deal with imbalanced data that is typically encountered in network security problems \cite{ensemble_rusboosting}. \\
\indent Table \ref{labeled_dataset_results} illustrates the performance of the proposed optimized ensemble tree-based learning model in comparison to the aforementioned classification learning models. It can be seen that the proposed framework outperforms the other models in terms of accuracy and precision. More specifically, it achieved an additional $\sim$1\% improvement in both accuracy and precision. Moreover, it also has comparable performance in terms of recall/sensitivity which is illustrated by the better F-score measure. This shows that the proposed framework is effective in detecting intrusions and network attacks in IIoT devices and environments. 
\begin{table}[!ht]
	\centering
	\caption{Performance Evaluation of Base and Optimized Classifiers}
	\scalebox{0.85}{
		\begin{tabular}{|p{2.7cm}|p{1.4cm}|p{1.4cm}|p{1.4cm}|p{1.1cm}|}	\hline
			Algorithm & Accuracy (\%) & Precision (\%)& Recall (\%)&F-score\\ \hline
			DT \cite{Moubayed_DNS2}&97.9&95.8&97.4&0.966\\ \hline
			Optimized DT &98.1&95.4&98.3&0.969\\ \hline
			Bagging Ensemble Trees\cite{Moubayed_DNS3}&98.2&96.1&98.1&0.971\\ \hline
			Boosting Ensemble Trees\cite{Moubayed_DNS3}&98.2&95.3&\textbf{99.1}&0.971\\ \hline
			RUSBoosting Ensemble Trees&97.9&94.5&99.1&0.967\\ \hline
			Optimized Ensemble Trees&\textbf{98.4}&\textbf{96.1}&98.7&\textbf{0.973}\\ \hline
		\end{tabular}
	}
	\label{labeled_dataset_results}
\end{table}

\indent To further illustrate the impact of the BO-GP optimization process on the performance of the learning models, Fig. \ref{Min_class_error} plots the minimum classification error progression during the optimization process for both the optimized DT and the proposed optimized ensemble learning models. It can be seen that the optimization process was successful in reducing the classification error for both models in a relatively short number of iterations (within 30 iterations). It can also be seen that the optimized ensemble learning model has a lower classification error which is also reflected in the detection accuracy, precision, recall, and F-score measures. This is because this model combines the benefits of the ensemble process and the hyper-parameter optimization process. This again emphasizes the need for and the positive impact of optimizing the learning models' parameters to improve their performance. Additionally, the optimized ensemble model reached its minimum error in less iterations (22 iterations) compared to the optimized DT model (28 iterations). This illustrates the improved efficiency of the proposed model compared to other optimized learning models previously proposed in the literature.
\section{Conclusion \& Future Works}\label{conc}
\indent The continued growth in the deployment of Internet-of-Things (IoT) devices has been fueled by the increased connectivity demand, particularly in industrial environments. However, this has led to an increase in the number of network related attacks due to the increased number of potential attack surfaces. Among the most vulnerable IoT devices are the industrial IoT (IIoT) ones. IIoT devices are prone to various network related attacks that can have severe consequences on the manufacturing process as well as on the safety of the workers in the manufacturing plant. Although researchers have proposed multiple potential mechanisms to protect such environments, there stills needs to have more effective IIoT attack detection mechanisms to address such concerns.\\ 
\indent To address this issue, this paper proposes a machine learning (ML) based framework by combining Bayesian Optimization-Gaussian Process (BO-GP) with an ensemble tree-based learning model to improve the performance of intrusion and attack detection in industrial IoT environments. Experimental results showed that the detection accuracy, precision, and F-score improved when compared to the standard tree and ensemble tree models. This highlighted the effectiveness of the proposed framework in detecting intrusions in IIoT environments.\\
\indent To extend this work, several directions can be explored. One such direction is to investigate the performance of other optimized learning models including deep learning models. A second direction is exploring the time series-based learning models that can uncover temporal patterns and investigating their impact on the intrusion detection performance.

\small
\bibliographystyle{IEEEtran}
\bibliography{IIoT_Ref}

\end{document}